# Factorization-based Lossless Compression of Inverted Indices

George Beskales [*]
University of Waterloo
Waterloo, ON, Canada
gbeskale@cs.uwaterloo.ca

Marcus Fontoura [†]
Google Inc.
Mountain View, CA, USA
marcusf@google.com

Maxim Gurevich
Yahoo! Labs
Santa Clara, CA, USA
maximg@yahoo-inc.com

Sergei Vassilvitskii
Yahoo! Labs
Santa Clara, CA, USA
sergei@yahoo-inc.com

Vanja Josifovski
Yahoo! Labs
Santa Clara, CA, USA
vanjaj@yahoo-inc.com

## ABSTRACT

Many large-scale Web applications that require ranked top-$k$ retrieval such as Web search and online advertising are implemented using inverted indices. An inverted index represents a sparse term-document matrix, where non-zero elements indicate the strength of term-document association. In this work, we present an approach for lossless compression of inverted indices. Our approach maps terms in a document corpus to a new term space in order to reduce the number of non-zero elements in the term-document matrix, resulting in a more compact inverted index. We formulate the problem of selecting a new term space that minimizes the resulting index size as a matrix factorization problem, and prove that finding the optimal factorization is an NP-hard problem. We develop a greedy algorithm for finding an approximate solution.

A side effect of our approach is increasing the number of terms in the index, which may negatively affect query evaluation performance. To eliminate such effect, we develop a methodology for modifying query evaluation algorithms by exploiting specific properties of our compression approach.

Our experimental evaluation demonstrates that our approach achieves an index size reduction of 20%, while maintaining the same query response times. Higher compression ratios up to 35% are achievable, however at the cost of slightly longer query response times. Furthermore, combining our approach with other lossless compression techniques, namely variable-byte encoding, leads to index size reduction of up to 50%.

## 1. INTRODUCTION

Web search engines and other large-scale information retrieval (IR) systems typically have to process query workloads of thousands of requests per second over large collections of documents. Usually, the result of the retrieval is a ranked list of the top few ($k$) results. Top-$k$ evaluation of textual queries is used in a large number of Web applications such as search, textual advertising, and product recommendation.

Top-$k$ retrieval can be defined as follows. Given a query $Q$ and a document corpus $Docs$, find the $k$ documents $\{D_1, D_2, \ldots, D_k\} \subset Docs$ that have the highest score, according to some scoring function $Score(D, Q)$. Both the query and the documents are sets of *terms* from the same high-dimension space. Scoring is usually performed based on the overlapping terms between the query and the document (i.e., the intersection between the document and the query terms). The document corpus $Docs$ can be represented as a two dimensional matrix, denoted as $V$, with $m$ terms and $n$ documents (Figure 1). In general, the values of elements in $V$ measure how strongly the terms are associated with the corresponding documents. For example, one measure is the term frequency, which is the number of occurrences of a term in a document. Given an example query $Q$ in Figure 1, the shaded portion of the matrix is used for its evaluation.

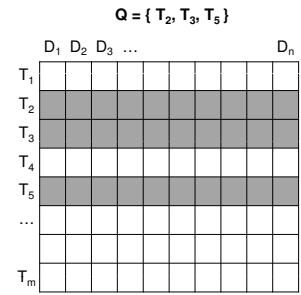

**Figure 1: Document corpus as a matrix $V$**

*Inverted indices* are the prevailing implementation of scalable top-$k$ retrieval. In an inverted index, each term $T$ appearing in the corpus $Docs$ is associated with a *posting list*, which enumerates the documents that contain $T$. An inverted index is a sparse representation of the matrix $V$ that stores only non-zero matrix elements.

In several applications, top-$k$ queries are processed while the user is waiting for the reply, which imposes very strict bounds on query latency. Due to such requirements, memory-resident indices are becoming more popular in current search engines. To lower the amount of required memory, and hence the system cost, compression techniques (e.g., [5, 6, 22, 25, 26, 27]) are heavily used to reduce the size of the inverted indices. Compression techniques are mainly divided into two categories: lossless compression, where quality of results are not affected by the compression, and lossy compression, where results quality might be affected. Lossy techniques typically trade index size for retrieval accuracy [5, 6], while lossless techniques exploit the properties of the document corpus for compactly encoding information in individual posting lists, such as documents identifiers [27, 26], and term positions [22, 25].

In this paper, we propose a novel lossless compression technique that holistically compresses multiple posting lists by taking advan-tage of similarities between them. This type of compression can be applied before the standard per-posting list compression in order to combine the benefits of both methods. In spirit, the technique presented in this paper is related to matrix factorization methods such as Non-negative Matrix Factorization (NMF) [13, 16], Latent Dirichlet Allocation [3], Singular Value Decomposition [23], and Principal Component Analysis [14], among others. All of these techniques map the documents from the space determined by the original terms into a lower dimensional space. However, unlike previous factorization methods, we aim at providing an *exact* factorization of the input matrix in order to avoid any information loss, while reducing the number of non-zero elements in the resulting factors. Furthermore, we do not restrict the new space to have a small number of terms.

---
[*] Work done while interning at Yahoo! Labs
[†] Work done while affiliated with Yahoo! Labs

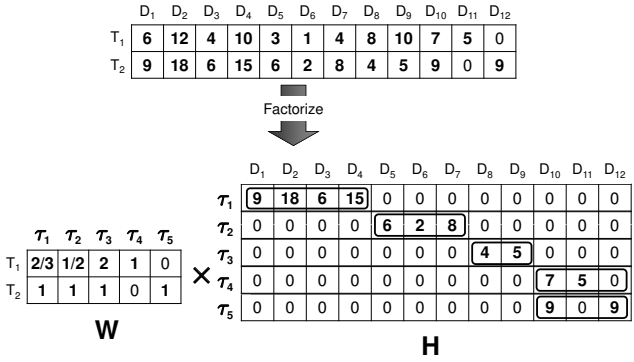

**Figure 2: Matrix representation for combining two terms**

For example, the top matrix in Figure 2 is factored into two matrices such that: (1) the product of the factors is equal to the input matrix, and (2) the factor matrices contain fewer non-zeros than the input matrix. Note that the rank of the second factor (five), is higher than the rank of the input matrix (two). To answer user queries, we use the first factor to map (i.e., rewrite) query terms from the original term space $T$ to a new space $\tau$ of meta-terms (e.g., $T_1$ is mapped to $\tau_1, \tau_2, \tau_3, \tau_4$ in Figure 2). The rewritten query is used for searching the second factor, which represents the compressed inverted index, using any top-$k$ search algorithms.

In this paper, we prove that finding the optimal (i.e., the sparsest) factorization is NP-hard. We develop a greedy algorithm that efficiently finds an approximate solution. The core of the algorithm is: (1) efficiently identifying segments of various posting lists that are identical up to a multiplicative factor, (2) extracting a copy of such common segments representing new meta-terms, and (3) removing the common segments from the original lists. Term mappings are constructed such that the original terms are mapped to both the updated original lists (with the common segment removed) and to the newly created meta-terms.

Although lossless compression helps reducing the amount of memory required for storing a given index, it usually incurs a computational overhead due to the need of decompressing the index data [22, 26]. Such overhead should be minimized in order to keep query latency small. The computational overhead in our approach is due to rewriting a query using a number of meta-terms that is greater than the number of the original terms. For example, the number of meta-terms in Figure 2 is five, while the original number of terms is two. We show how to eliminate such negative effect by exploiting some unique characteristics of our compression technique. More specifically, we show that standard query processing approaches such as No-Random-Access (NRA) algorithm [10] can be modified to search the compressed index as fast as the original algorithm searches the uncompressed index.

We evaluate the proposed compression techniques on TREC WT10g dataset [1]. The experiments show that our compression algorithm reduces the index size by up to 35%. Furthermore, integrating our approach with a standard lossless compression technique, namely variable-byte encoding [19], pushes the space savings to 50%. We show that moderate compression (e.g., 20%) incurs no overhead on query evaluation performance, while higher compression ratios incur a negligible overhead.

In summary, the contributions of the paper are as follows.

- We propose a novel approach to lossless compression of inverted indices that is based on exact matrix factorization. We prove that obtaining the optimal factorization in NP-hard.

- We propose a greedy algorithm for exact matrix factorization, and show how to parallelize our algorithm using the MapReduce paradigm. We show that it is still possible to incrementally update compressed indices with a minimal cost.

- We demonstrate how to eliminate query evaluation overhead due to decompression by exploiting characteristics of the compressed index.

- We experimentally evaluate our techniques on the standard TREC WT10g dataset.

The remainder of the paper is organized as follows. In Section 2, we introduce basic concepts and notation used throughout the paper. In Section 3, we establish the link between index compression and matrix factorization. Section 4 describes the proposed factorization algorithm, and how to update a compressed index to accommodate new documents. In Section 5, we show how to modify search algorithms to reduce query response time. The experimental evaluation is presented in Section 6. We discuss the related work in Section 7. Finally, we conclude the paper in Section 8.

## 2. PRELIMINARIES

In this paper, we use the vector-space representation of documents and queries. That is, documents and queries are represented as vectors in a multidimensional space where each term is a dimension. Let $\Omega = \{T_1, T_2, \ldots, T_m\}$ be a set of terms. A document $D_j$ is a vector $(d_j^1, d_j^2, \ldots, d_j^m)$. When a term $T_i$ occurs in a document $D_j$, the element $d_j^i$ is non-zero, and its value is typically related to the number of times $T_i$ occurs in the document. Similarly, a query $Q$ is represented by a vector $(w_1, w_2, \ldots, w_m)$, where non-zero elements correspond to terms appearing in the query, and their values are term weights in the query.

Given a document corpus of $n$ documents $Docs = \{D_1, D_2, \ldots, D_n\}$ and a query $Q$, a common task in many information retrieval systems is to retrieve the $k$ documents with the highest score according to some scoring function $Score(D, Q)$. In this work, we assume the scoring function is defined as the inner product of document and query vectors. That is,

$$Score(D_j, Q) = \sum_{i=1}^{m} d_j^i \cdot w_i. \qquad (1)$$

Many information retrieval systems use inverted indices as their main data structure for top-$k$ retrieval. An inverted index is a collection of *posting lists* $L_1, L_2, \ldots, L_m$: a list for each term in $\Omega$. List $L_i$ is a vector containing weights of term $T_i$ in all documents (i.e., $L_i = (d_1^i, d_2^i, \ldots, d_n^i)$).

For typical document collections, the majority of the values in posting lists are zeros. Thus, inverted indices use sparse representation, where zero entries are omitted. Specifically, each posting list $L_i$ contains postings of the form $\langle \textit{docID, payload} \rangle$, where *docID* is the document identifier $D_j$, and the *payload* contains the (non-zero) value $d_j^i$.

Given a query $Q$, top-$k$ search algorithms use posting lists of terms that have non-zero weights in $Q$ to obtain the top-$k$ documents. A naïve algorithm would examine all entries in the relevant posting lists, compute the scores of found documents, and return the top-$k$ documents. However, the total number of documents in the relevant posting lists is typically much larger than $k$, especially when $Q$ contains frequent terms. Many top-$k$ search algorithms (e.g., [4, 10]) aim at retrieving the top-$k$ documents while examining only a fraction of entries in the relevant posting lists.

REMARK In this work, we assume that the payload does not contain additional information such as term position, typesetting, etc. Primitive payloads (i.e., consisting of only term frequencies) can be found in many large-scale applications such as computational advertising, where documents are relatively short, and extra information such as term positions are not informative. Moreover, some applications employ a two-tier retrieval. That is, candidate results are first obtained from an inverted index consisting of primitive information only. In the second phase, candidate results are re-scored using a more accurate function that considers additional information. The additional information is typically fetched from a forward index, and does not need to be stored in the inverted index.

## 3. INDEX COMPRESSION AS MATRIX FACTORIZATION

We represent an index by an $m \times n$ term-document matrix $V$, where rows are posting lists and columns are document vectors (Figure 1). We denote by $V[T, D]$ the value of the element in $V$ corresponding to a row $T$ and a column $D$. We use $\|V\|_0$ to denote the number of non-zero elements in $V$. Top-$k$ retrieval corresponds to computing a score vector $S$ that is equal to the product of a query vector $Q$ and the matrix $V$, and picking the top-$k$ documents with highest scores. Formally, $S^T = Q^T V$, where the superscript $T$ denotes the transpose operator.

Typical document collections, such as Web corpora, contain redundant elements in their term-document matrices due to duplicate or near-duplicate contents. For example, news articles are usually shared across multiple Web sites. In this case, two documents $D_x$ and $D_y$ that refer to the same article would contain several identical sentences consisting of terms $T_a, T_b, \ldots, T_p$. Consequently, the term-document matrix would contain two identical sets of values: $V[T_a, D_x], \ldots, V[T_p, D_x]$, and $V[T_a, D_y], \ldots, V[T_p, D_y]$. Another example that leads to redundancy in term-document matrix is co-occurrence of subsets of terms in multiple documents. For example, terms "Britney" and "Spears" usually co-occur in documents related to music. In this case, the term-document matrix will contain two identical sets of values: $V[T_x, D_a], \ldots, V[T_x, D_p]$, and $V[T_y, D_a], \ldots, V[T_y, D_p]$, where $T_x$ and $T_y$ are co-occurring terms in a set of documents $\{D_1, \ldots, D_p\}$.

A known technique for reducing redundancy in a matrix is matrix factorization. The simplest form of factorization is decomposing $V$ into two matrices: an $m \times r$ matrix $W$ and an $r \times n$ matrix $H$, such that $V = WH$. Note that since our goal is lossless index compression, we consider the exact formulation and not the approximate one ($V \approx WH$). In our case, the objective function is to minimize the total number of non-zero elements in $W$ and $H$ (i.e., $\|W\|_0 + \|H\|_0$).

Intuitively, factoring $V$ into $WH$ transforms the set of terms $\Omega$ into another space, denoted $\Theta$, consisting of $r$ *meta-terms* $\{\tau_1, \tau_2, \ldots, \tau_r\}$. Matrix $W$ linearly maps terms in $\Omega$ to meta-terms in $\Theta$ (and vice-versa), while matrix $H$ represents the inverted index of $Docs$ in the space of meta-terms. Figure 2 is an illustration of such a factorization, where terms $\{T_1, T_2\}$ are linearly mapped into meta-terms $\{\tau_1, \ldots, \tau_5\}$ using matrix $W$, and documents are represented as combinations of these meta-terms in matrix $H$. Note that although $r > m$ (i.e., the number of rows in $H$ are greater than the number of rows in $V$), the number of non-zeros in $W$ and $H$ is less than the number of non-zeros in $V$.

Evaluation of query $Q$ is performed on the inverted index represented by $H$, after rewriting $Q$ according to $W$. Specifically, we rewrite the query vector $Q$ into vector $Q'$ such that $Q'^T = Q^T W$. In other words, each term $T$ with non-zero weight in $Q$ is replaced by a set of meta-terms $\{\tau : W[T, \tau] \neq 0\}$. The weight of each term $\tau$ in $Q'$ is $w \cdot W[T, \tau]$, where $w$ is the weight of $T$ in $Q$. Once $Q$ is rewritten into $Q'$, any standard search algorithm can be used to retrieve the top-$k$ documents from the compressed index $H$ using query $Q'$. The following theorem proves that searching the original inverted index using $Q$ is equivalent to searching the index represented by $H$ using $Q'$.

THEOREM 1. *Let $W$ and $H$ be the result of factoring $V$ (i.e., $WH = V$). Let $\mathcal{A}(V, Q, k)$ be the top-k documents for query $Q$ using inverted index $V$ and the scoring function in Equation 1 (ties in scores are broken by some predefined criteria). Let the rewritten query be $Q'$ such that $Q'^T = Q^T W$. Then, $\mathcal{A}(V, Q, k) = \mathcal{A}(H, Q', k)$.*

PROOF. Let $Score(D_j, Q, V)$ denote the score of a document $D_j$, given a query $Q$ and an inverted index represented by matrix $V$. Since the top-k results are selected based on document scores computed using Equation 1, we only need to show that $Score(D_j, Q, V) = Score(D_j, Q', H)$ for all $D_j$. The value of $Score(D_j, Q, V)$ can be rewritten as the dot product $Q \cdot V[:, D_j]$, where $V[:, D_j]$ denotes the vector corresponding to the column $D_j$ in $V$. Then,

$$\begin{aligned} Score(D_j, Q, V) &= Q \cdot V[:, D_j] = Q^T V[:, D_j] \\ &= Q^T WH[:, D_j] = Q'^T H[:, D_j] \\ &= Q' \cdot H[:, D_j] = Score(D_j, Q', H). \end{aligned}$$

□

An immediate consequence of Theorem 1 is that standard top-$k$ algorithms can still be used for searching the compressed indices without any loss in precision or recall.

Typically, inverted indices contain non-negative term-document weights, e.g., reflecting term frequencies. The non-negativity of weights is exploited in some top-$k$ retrieval algorithms such as WAND [4]. In order to be able to use such algorithms over compressed indices, we aim at preserving the non-negativity of $V$ in the factor matrices $W$ and $H$.

REMARK It is possible to interpret the intermediate space $\Theta$ as a space of *meta-documents*, rather than *meta-terms*. In this case, the matrix $W$ represents an inverted index of meta-documents in the original term space $\Omega$, and $H$ is a mapping from meta-documents to documents in $Docs$. However, under such interpretation, existing top-$k$ retrieval algorithms that employ early termination cannot be used on $W$, since top-$k$ meta-documents do not necessarily contain the top-$k$ documents. For example, in Figure 2, suppose that the entities $\tau_1, \ldots, \tau_5$ represent a set of meta-documents, and suppose we want to compute top-1 document for query $Q = (1, 0)$, which is $D_2$. Applying the top-$k$ algorithm to $W$ returns $\tau_3$, which is then mapped to the document set $\{D_8, D_9\}$ that does not include the correct top-1 result $D_2$.

## 4. SPARSE MATRIX FACTORIZATION

In this section, we consider the following exact sparse matrix factorization problem. Given a matrix $V$, obtain $W$ and $H$ that

$$\text{minimize} \quad \|W\|_0 + \|H\|_0$$
$$\text{subject to} \quad WH = V$$

Unlike typical factorization problems, we do not impose any restriction on the dimensionality $r$ of the intermediate space (i.e., $|\Theta|$). In particular, it is allowed to be higher than $\max(n, m)$. Requiring $r$ to be much lower than the original dimensions $n$ and $m$ prohibits sparse and exact solutions, which is required in our problem. Thus, existing factorization techniques are not appropriate (see more details in Section 7), and we resort to developing a new factorization approach.

The following theorem states that the problem of obtaining the sparsest exact factorization is NP-hard.

THEOREM 2. *Given a matrix $V$, the problem of obtaining two matrices $W$ and $H$, subject to the constraint $WH = V$, such that $\|W\|_0 + \|H\|_0$ is minimum is NP-hard.*

PROOF. We prove the claim by reduction from the NP-complete SPARSESTVECTOR problem [11]: given a full rank $m \times n$ matrix $A$, and an $m \times 1$ vector $b$, find an $n \times 1$ vector $x$ with minimal $\|x\|_0$ such that $Ax = b$. Given an instance $(A, b)$ of the SPARSESTVECTOR, we construct an instance of the sparse matrix factorization problem as follows. Let $V$ be a matrix obtained by concatenating $A$, $p = n(m + 1) + 1$ times horizontally, followed by the vector $b$: $V = [A\ A\ \ldots\ A\ b]$. Note that $V$ is an $m \times (np + 1)$ matrix.

One solution is to factor $V = AB$ with $B = [I\ I \ldots\ I x]$, where $I$ is the $n \times n$ identity matrix, and $x$ is the optimum solution to the SPARSESTVECTOR problem. For this solution, $\|B\|_0 = np + \|x\|_0$, and the total solution cost $\|A\|_0 + \|B\|_0 = \|A\|_0 + np + \|x\|_0 \leq n(m + p + 1)$ .

Consider any other solution for factoring $V = WH$ such that $W$ is a $m \times k$ matrix, and $H$ is a $k \times (np + 1)$ matrix. Let $H = [H_1\ H_2 \ldots H_p\ y]$. In any optimal solution $\|H_1\|_0 = \|H_2\|_0 = \ldots = \|H_p\|_0$. Let us assume otherwise, i.e., there are indices $i$ and $j$ with $\|H_i\|_0 < \|H_j\|_0$. Let $H' = [H_1\ \ldots\ H_{j-1}\ H_i\ H_{j+1}\ \ldots\ H_p\ y]$, then $V = WH'$ and $\|H'\|_0 < \|H\|_0$, a contradiction with optimality of $(W, H)$.

Let $q = \|H_i\|_0$, for $i \in \{1, \ldots, p\}$. Assume that $q \geq (n + 1)$, therefore the cost of the solution is at least:

$$qp = (n+1)p = (n+1)(n(m+1)+1) = n^2m + nm + n^2 + 2n + 1.$$

But the cost of the $(A, B)$ solution is no more than $n(m+p+1) = n(m+n(m+1)+2) = n(m+nm+n+1) = nm+n^2m+n^2+2n$. Therefore the presented a solution is no longer optimal, which is a contradiction. Thus, for any optimal solution, $q < n + 1$.

Observe that since $A$ is full rank, $q \geq n$. Therefore, in any optimal solution $q = n$ and $H_i$ is a permutation of the identity matrix, for $i \in \{1, \ldots, p\}$. Therefore, $y$ is a permutation of the solution to the sparsest vector problem.

☐

Since obtaining the optimal factorization is computationally infeasible, we propose an iterative greedy algorithm for getting an exact factorization that might not have the minimal number of non-zeros. The key idea is to start with a trivial factorization $W_0 = I_m$ ($I_m$ is the identity matrix of rank $m$) and $H_0 = V$, and iteratively improve the current solution $(W_t, H_t)$ by a sequence of local transformations on $W_t$ and $H_t$, obtaining $W_{t+1}$ and $H_{t+1}$. Each step is guaranteed to reduce $\|W_t\|_0 + \|H_t\|_0$ while preserving two invariants: (1) $W_t H_t = V$, and (2) $\|W_{t+1}\|_0 + \|H_{t+1}\|_0 < \|W_t\|_0 + \|H_t\|_0$. Although our iterative algorithm does not necessarily reach an optimal solution, it achieves significant compression ratios after a few iterations (see Section 6).

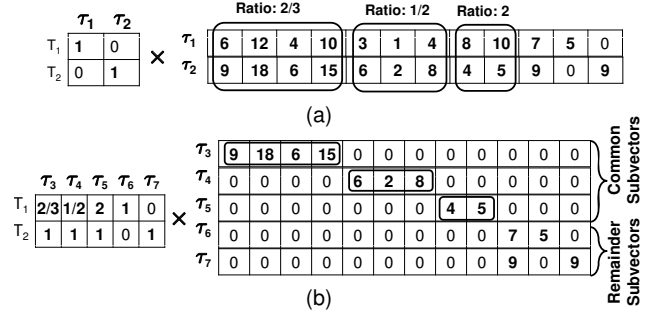

**Figure 3: Combining term vectors (a) matrices $W_t$ and $H_t$ (b) matrices $W_{t+1}$ and $H_{t+1}$ after combining $\tau_1$ and $\tau_2$**

The transformation performed at each step is based on the observation that correlated terms and documents induce correlated values in columns and rows of $V$. At step $t$, given matrices $W_t$ and $H_t$, the algorithm looks for a submatrix $H_t^s$ of $H_t$ defined by a subset $R$ of $H_t$'s rows, and a subset $C$ of $H_t$'s columns, such that the rank of the submatrix is one. That is, all rows of $H_t^s$ are multiples of each others:

$$\forall (\tau_i, \tau_j) \in R \times R, \ \forall (D_p, D_q) \in C \times C \quad \left(\frac{V[\tau_i, D_p]}{V[\tau_j, D_p]} = \frac{V[\tau_i, D_q]}{V[\tau_j, D_q]}\right) \quad (2)$$

Clearly, keeping only one representative row from $H_t^s$ and encoding other rows in $H_t^s$ as multiples of the representative row would reduce the number of non-zero values in $H_t$. Unfortunately, identifying the largest submatrix $H_t^s$ is equivalent to the problem of finding the largest bi-cluster [18], which is an NP-hard problem. Thus, our algorithms considers only submatrices consisting of two rows (i.e., $|R| = 2$) at each step.

For efficiency, our algorithm identifies $z$ rank-1 submatrices at each iteration that are composed of two rows $R = \{\tau_i, \tau_j\}$ and $z$ sets of columns $C_1, \ldots, C_z$ (i.e., Equation 2 holds for submatrix $(R, C_1)$ through $(R, C_z)$). Then, rows $\tau_i$ and $\tau_j$ can be rewritten as linear combinations of a set of $z$ common subvectors, denoted $\tau_{r+1}, \ldots, \tau_{r+z}$, and two remainder vectors $\tau_{r+z+1}$ and $\tau_{r+z+2}$ that contain values of documents that are not in $C_1 \cup \cdots \cup C_z$. More specifically,

$$\tau_i = \alpha_1 \cdot \tau_{r+1} + \alpha_2 \cdot \tau_{r+2} + \cdots + \alpha_z \cdot \tau_{r+z} + \tau_{r+z+1} \quad (3)$$
$$\tau_j = \beta_1 \cdot \tau_{r+1} + \beta_2 \cdot \tau_{r+2} + \cdots + \beta_z \cdot \tau_{r+z} + \tau_{r+z+2} \quad (4)$$

Vectors $\tau_{r+1}, \ldots, \tau_{r+z+2}$ are appended to matrix $H_t$, and vectors $\tau_i$ and $\tau_j$ are removed from $H_t$, resulting in matrix $H_{t+1}$. Matrix $W_t$ is modified to map original terms to $\tau_{r+1}, \ldots, \tau_{r+z+2}$ instead of $\tau_i, \tau_j$, resulting in matrix $W_{t+1}$. Algorithm 1 describes the procedure in more details. Function $GetCorrelatedSubmatrices(H_t)$, which we describe in Sec-tion 4.1, is responsible for extracting the sets $R$ and $C_1, \ldots, C_z$ that maximize space saving.

Figure 3 shows an example of combining two meta-terms $\tau_1$ and $\tau_2$ into common subvectors $\tau_3, \tau_4, \tau_5$, and remainder vectors $\tau_6, \tau_7$. Without loss of generality, we assume hereafter that $\beta_1 = \cdots = \beta_z = 1$. We ensure that $\alpha_p \neq \alpha_q$ for $p \neq q$ (otherwise, we combine $\tau_{r+p}$ and $\tau_{r+q}$ into one subvector). An important consequence is that $C_1, \ldots, C_z$ are pairwise disjoint. We rely on this property to improve query evaluation performance (Section 5).

---

**Algorithm 1** `ComputeFactorization(V)`

1: $W_0 \leftarrow I_m; H_0 \leftarrow V$
2: $r \leftarrow m$
3: $t \leftarrow 0$
4: **repeat**
5:     $H_{t+1} \leftarrow H_t$
6:     $(R, C_1, \ldots, C_z) \leftarrow GetCorrelatedSubmatrices(H_t)$
7:     **if** failed to find correlated submatrix **then**
8:       **break**
9:     remove rows $R = \{\tau_i, \tau_j\}$ from $H_{t+1}$, and add the new rows $\tau_{r+1}, \ldots, \tau_{r+z+2}$ to $H_{t+1}$
10:    construct an $r \times (r+z+2)$ transformation matrix $W_M$ that linearly maps $\tau_i$ and $\tau_j$ to $\tau_{r+1}, \ldots, \tau_{r+z+2}$ using Equations 3 and 4, and trivially maps all other meta-terms in $W_t$ to themselves.
11:    $W_{t+1} \leftarrow W_t W_M$.
12:    $r \leftarrow r + z + 2$
13:    $t \leftarrow t + 1$
14: **until** $\|W_t\|_0 + \|H_t\|_0$ converges
15: **return** $W_t, H_t$

---

## 4.1 Identifying Correlated Submatrices

The goal of function $GetCorrelatedSubmatrices(H_t)$ is to return correlated submatrices, defined by $R = \{\tau_i, \tau_j\}$ and $C_1, \ldots, C_z$. Our algorithm heuristically finds the submatrices that would result in the highest reduction of space. In the following, we describe how to find the sets $C_1, \ldots, C_z$ given $R$, formulate the potential saving from combining two given meta-terms, and finally how to find $R$.

First, we show how to compute $C_1, \ldots, C_z$, given the two meta-terms $\tau_i$ and $\tau_j$ to combine. Denote by $\tau[p]$ the value of the element at index $p$ in a row $\tau$ in $H_t$. For two rows $\tau_i$ and $\tau_j$ in $H_t$, we compute a vector $\gamma$ of length $n$ as follows:

$$\gamma[q] = \begin{cases} \frac{\tau_i[q]}{\tau_j[q]} & \text{if } \tau_i[q] \neq 0 \text{ and } \tau_j[q] \neq 0 \\ 0 & \text{otherwise} \end{cases}$$

for $1 \leq q \leq n$.

Each set $C_p$ is a subset of documents (columns in $H_t$) that have the same non-zero value in $\gamma$. For example, in Figure 3, the first four cells have the same value in $\gamma$, namely $2/3$, and thus constitute a common subvector ($\tau_3$).

The space saving resulting from combining $\tau_i$ and $\tau_j$ is computed as follows. Combining $\tau_i$ and $\tau_j$ in Algorithm 1 reduces the number of non-zero elements in $H_t$ by $\sum_{p=1}^{z} |C_p|$ because each subvector $\tau_{r+p}$ corresponding to $C_p$ is stored twice in $H_t$ and only once in $H_{t+1}$. On the other hand, combining $\tau_i$ and $\tau_j$ means that all the terms in $W_t$ that were mapped to either $\tau_i$ or $\tau_j$ are now mapped to additional $z$ meta-terms (i.e., $\tau_{r+1}, \ldots, \tau_{r+z}$) in $W_{t+1}$, which increases the number of non-zero elements in $W_{t+1}$. For example, in Figure 3, $T_1$ is originally mapped to $\tau_1$. After combining $\tau_1$ and $\tau_2$, $T_1$ is mapped to extra 3 meta-terms, namely $\tau_3, \tau_4, \tau_5$, (besides the remainder meta-term $\tau_6$), which results in three additional elements in $W$. Formally, the overall space saving when combining $\tau_i$ and $\tau_j$ is:

$$saving(\tau_i, \tau_j, C_1, \ldots, C_z, W_t) = \sum_{p=1}^{z} |C_p| \qquad (5)$$
$$- z \cdot |\{T \in \Omega : W_t[T, \tau_i] \neq 0 \vee W_t[T, \tau_j] \neq 0\}|$$

In the following, we describe how to efficiently identify a pair of rows in $H_t$, denoted by $R$, with the highest potential savings. A straightforward approach is to compute the potential space saving, based on Equation 5, for all pairs of rows and return the pair with the highest savings. Unfortunately, the complexity of such approach is quadratic in the number of rows in $H_t$, which is prohibitively expensive. To reduce the number of pair-wise comparisons, we use a *blocking* technique to prune a large number of pairs that have low potential space saving. Blocking techniques have frequently been used in clustering algorithms that rely on pairwise comparison (e.g., [20]). The main goal of a blocking technique is to partition the set of objects into multiple blocks such that "similar" objects are placed in the same block. Thus, we only need to compare pairs of objects that belong to the same block.

Recall that our distance metric for comparing two rows $\tau_i$ and $\tau_j$ is the potential reduction in space resulting from combining them. Let $\tau_i \cap \tau_j$ be the set of documents that have non-zero value in both $\tau_i$ and $\tau_j$ in $H_t$. The maximum possible savings can be obtained when all elements in $\tau_i \cap \tau_j$ are placed in the same common subvector. Therefore, we use the *overlap* between rows (i.e., $|\tau_i \cap \tau_j|$) as an upper bound of the potential savings. We thus place the rows with high overlap in the same block.

Since computation of overlap is expensive because of the large vector lengths, we approximate it using sketching. We partition documents $\{D_1, \ldots, D_n\}$ into $\lambda$ disjoint groups, denoted $G_1, \ldots, G_\lambda$, by assigning each document to a randomly selected group. For each row $\tau_i$, we compute a $\lambda$-dimensional vector $S_i$ such that $S_i[p]$, $1 \leq p \leq \lambda$, is equal to the number of documents in $G_p$ that are associated to $\tau_i$ in $H_t$. Vector $S_i$ is the sketch of $\tau_i$. The blocking algorithm picks the dimension $dim \in \{1, \ldots, \lambda\}$ with the largest variance across sketches $S_1, \ldots, S_r$. Dimension $dim$ is used for splitting the rows into two blocks such that the first (respectively, second) block contains rows with value of dimension $dim$ below (respectively, above) the median. The algorithm recursively applies the same process until block sizes are smaller than a predefined threshold $B$. We experimentally analyze the effect of parameters $\lambda$ and $B$ in Section 6. Once sufficiently small blocks are identified, we find a pair of rows that maximizes space savings by brute-force computation in each block.

## 4.2 Limiting the Number of Subvectors

Recall that our compression approach iteratively reduces the index size at the cost of increasing the number of meta-terms. That is, there is a trade-off between the space savings and the increase in the number of meta-terms. In particular, it may not be worthwhile to introduce new meta-terms whose space savings are below some threshold. We note that the length of a new meta-term $\tau_{r+p}$ represents its maximum potential saving, according to Equation 5. Therefore, we modify algorithm $GetCorrelatedSubmatrices(H_t)$ such that it generate a new meta-term only if its length is greater than or equal to a threshold $\mu$. Consider the example depicted in Figure 3 and let $\mu = 3$. Then, only the first two meta-terms $\tau_3, \tau_4$ would be generated, while the third meta-term $\tau_5$ would not be generated (i.e., it becomes part of the remainder meta-terms $\tau_6, \tau_7$). In Section 6, we experimentally analyze the effect of $\mu$ on the number of meta-terms and the compression ratio.

### 4.3 MapReduce Implementation

Even moderate inverted indices consist of millions of documents and terms, making sequential implementation of our iterative algorithm impractical. To scale the algorithm to large matrices, we parallelize it according to the MapReduce model [15].

In each iteration of Algorithm 1, we combine two rows of matrix $H_t$, generating $z + 2$ new rows in matrix $H_{t+1}$ and updating values in matrix $W_{t+1}$. These operations are performed independently of other rows in $H_t$. Moreover, the rows and values written to $H_{t+1}$ and $W_{t+1}$ depend *only* on the rows being combined. These observations allow the following parallelization scheme: (1) identify several disjoint pairs of rows, (2) combine all pairs in parallel and emit the new meta-terms and their coefficients, and construct $H_{t+1}$ and $W_{t+1}$. Algorithm 2 describes the details of the procedure.

Since it is impractical to run the algorithm until full convergence, we use a parameter $\delta$ to control the number of iterations. Once the space savings resulting from the current iteration is below $\delta$, the algorithm stops and returns the current matrices $W_t$ and $H_t$. Function $GetCorrelatedRowsMR(H_t)$ obtains a set of independent row pairs $\{(\tau_i, \tau_j), \ldots\}$ in $H_t$ to combine, such that the overall savings is maximized. The function computes sketches of the rows in $H_t$ in parallel (map phase), and then partition the rows into blocks according to sketches (reduce phase). In each block, the potential space savings from combining each pair of rows is computed, and independent (i.e., disjoint) pairs with the highest savings are selected using a technique from [12].

---

**Algorithm 2** `ComputeFactorizationMR(V)`
1: $W_0 \leftarrow I_m$
2: $H_0 \leftarrow V$
3: $r \leftarrow m$
4: $t \leftarrow 0$
5: **repeat**
6:    $\{(\tau_i, \tau_j), \ldots\} \leftarrow GetCorrelatedRowsMR(H_t)$
7:    $(H_{t+1}, W_{t+1}) \leftarrow CombineMR(\{(\tau_i, \tau_j), \ldots\}, W_t, H_t)$
8:    $t \leftarrow t + 1$
9: **until** $\frac{(\|W_{t-1}\|_0 + \|H_{t-1}\|_0) - (\|W_t\|_0 + \|H_t\|_0)}{(\|W_{t-1}\|_0 + \|H_{t-1}\|_0)} < \delta$
10: **return** $W_t, H_t$

---

Function $CombineMR(\{(\tau_i, \tau_j), \ldots\}, W_t, H_t)$ computes matrix $H_{t+1}$ by combining pairs of terms that are obtained by $GetCorrelatedRowsMR(H_t)$. The set of row pairs $\{(\tau_i, \tau_j), \ldots\}$ is cached at each mapper/reducer. A map task associates each row in $H_t$ to a key referring to the row pair it belongs to (or to itself if it is not part of any pair). Then, rows that belong to the same row pair are grouped at the reduce task, and the resulting rows of $H_{t+1}$ are computed. A term-transformation matrix $W_M$ is computed analogously. Finally, the matrix $W_{t+1}$ is computed as the product $W_t W_M$ by caching $W_M$ at all mappers/reducers that process rows of $W_t$ and outputs the rows of $W_{t+1}$. Based on the overall space reduction and the parameter $\delta$, the algorithm decides whether to start a new iteration or to terminate.

### 4.4 Updating the Compressed Index

Document corpora that are extracted from the Web are frequently updated due to the constant flow of new documents, removing obsolete documents, and modifying existing documents. The frequency of such updates requires the ability to update inverted indices incrementally, without rebuilding the entire index on each update. In the following, we describe how to incrementally update factorized indices. We focus on two operations: adding new documents and removing existing documents. Updating an existing document can be implemented by removing the old version of the document and adding the new version to the index.

In regular inverted indices, adding a new document is implemented by assigning the document a new document identifier and inserting a posting into the posting list of each term that appears in the document. The insertion position in the posting list depends on how posting lists are ordered.

Recall that our compression approach maps each term $T_i \in \Omega$ to a set of meta-terms, denoted $M(T_i) \subseteq \Theta$, through the matrix $W$. At least one meta-term in $M(T_i)$ is a remainder meta-term, denoted $Rem(T_i)$, that is uniquely mapped to $T_i$. That is, $W[T_i, Rem(T_i)] = 1$, and $\forall j \neq i(W[T_j, Rem(T_i)] = 0)$. It is straightforward maintain a mapping $T_i \rightarrow Rem(T_i)$ during the compression procedure. In order to add a new document $D$ that mention term $T_i$, it is sufficient to add a new posting to the posting list of $Rem(T_i)$. Thus, it is possible to accommodate frequent insertions of new documents through maintenance of remainder meta-terms only. Note that adding postings to remainder meta-terms does not allow the maximum possible space saving that can be achieved by rebuilding the compressed index from scratch. The reason is that redundancy in newly inserted documents is ignored. It is possible to reduce the overhead of a full index rebuild by continuing the compression algorithm from the current matrices $W$ and $H$ rather than starting with $W_0 = I_m$ and $H_0 = V$ (lines 1 and 2 in Algorithm 1).

Removing documents from the compressed index is achieved by removing all postings in the compressed index that refer to the removed document. This is equivalent to removing the entire column in $H$ that corresponds to the removed document.

## 5. OPTIMIZING QUERY PROCESSING

A possible side-effect of our compression approach is having a number of meta-terms in the rewritten query that is larger than the number of terms in the original query. Such increase can be quite significant as we demonstrate in Section 6.2, and can lead to a noticeable increase in query evaluation time.

In this section, we show how to mitigate this undesirable effect by exploiting unique characteristics of our compression scheme to improve the efficiency of typical top-k query processing algorithms. As a case study, we show how to modify the Non-Random-Access algorithm (NRA) [10]. Note that there exist a plethora of search algorithm that might be more efficient than NRA, especially for memory-resident indices. However, we chose the NRA algorithm mainly because of its simplicity to describe and analyze. The observations in this section can be exploited to adapt other search algorithms, provided that they use similar primitives to access posting lists.

We denote by $L_1, \ldots, L_h$ the posting lists corresponding to the terms with non-zero weight in query $Q$, where $h = \|Q\|_0$, and let $w_1, \ldots, w_h$ be the weights associated with $L_1, \ldots, L_h$ in $Q$. The score of a document $D$ can be rewritten as follows:

$$Score(D, Q) = \sum_{i=1}^{h} w_i \cdot L_i(D) \qquad (6)$$

where $L_i(D)$ denotes the weight of document $D$ in list $L_i$. The NRA algorithm requires posting lists to be sorted in descending order of document weights. We assume that weights of the query terms $w_1, \ldots, w_h$ are positive and that the number of documents in the corpus is greater than the number of documents to retrieve ($k$).

The NRA algorithm retrieves documents from lists $L_1, \ldots, L_h$ in a round robin order. The key insight that allows early termination is that having the lists sorted enable computing upper and lower score bounds for document scores. Every time a document is retrieved, the lower and upper bounds of retrieved documents, as well as unseen documents, are updated. Once there exist $k$ documents whose lower bounds are greater than or equal to the upper bounds of all other documents (including both seen and unseen documents), the algorithm terminates.

Score bounds are computed in the NRA algorithm as follows. Let $\overline{x}_i$ denote the weight of the last document retrieved from list $L_i$, if $L_i$ is not completely read by the algorithm, or 0 otherwise. During execution of the algorithm, the score upper bound of each retrieved document $D$, denoted $\overline{Score}(D, Q)$, is computed as follows:

$$\overline{Score}(D, Q) = \sum_{i=1}^{h} w_i \cdot \overline{L_i}(D). \quad (7)$$

where $\overline{L_i}(D)$ denotes the weight of $D$ in list $L_i$, if $D$ has appeared in $L_i$, and $\overline{x}_i$ otherwise. The upper bound for unseen documents is $\sum_{i=1}^{h} w_i \cdot \overline{x}_i$. Similarly, the lower bound of each retrieved document $D$, denoted $\underline{Score}(D, Q)$, is:

$$\underline{Score}(D, Q) = \sum_{i=1}^{h} w_i \cdot \underline{L_i}(D) \quad (8)$$

where $\underline{L_i}(D)$ denotes the weight of $D$ in list $L_i$, if $D$ has appeared in $L_i$, and 0 otherwise.

In the following, we describe our modifications to the NRA algorithm. Observe that the score upper bound of each retrieved document $D$ is computed by assuming that each undiscovered weight $L_i(D)$ is equal to $\overline{x}_i$ (Equation 7). Recall that all meta-term lists corresponding to the same original term are disjoint (Section 4). Thus, it is possible to compute a tighter score upper bound by setting undiscovered weight $L_i(D)$ to zero, instead of $\overline{x}_i$, if $D$ has appeared in any list $L_j$ such that $L_i$ and $L_j$ are disjoint.

Therefore, instead of considering lists of meta-terms independently, we create a two-level document retrieval scheme as follows. We create a *virtual list* for each original query term $T_i$. Each virtual list is traversed by probing the *disjoint* lists of the corresponding meta-terms. We use a priority queue $PQ_i$ to implement the virtual list of $T_i$.

Algorithms 3 and 4 describe how to initialize a priority queue and how to get next document, respectively. Let $M(T_i) \triangleq \{\tau : W[T_i, \tau] \neq 0\}$ be the set of meta-terms that term $T_i$ is rewritten into. Initialization of a priority queue $PQ_i$ is performed by inserting a pair $(\tau, D)$ for each meta-term $\tau$ in $M(T_i)$, where $D$ is the document with the highest score in $\tau$. The score of each pair $(\tau, D)$ in the queue is equal to the score of $D$ in list $\tau$ multiplied by the weight $W[T_i, \tau]$. Retrieving next document from the virtual list of $T_i$ is equivalent to retrieving the document in the pair $(\tau, D)$ at the head of the priority queue. After each retrieval from $PQ_i$, we insert a new pair $(\tau, D')$ in $PQ_i$, where $D'$ is the next document in $\tau$.

Modifying the NRA algorithm to use the virtual lists is straightforward: instead of initializing posting lists, the algorithm initializes priority queues $PQ_1, \ldots, PQ_h$ for the original query terms $T_1, \ldots, T_h$, and retrievals from each list $L_i$ are replaced by retrievals from the corresponding virtual list. The following theorem proves the correctness of the modifications, and gives an upper bound on the runtime overhead of the modified NRA algorithm.

**Algorithm 3** Initialize_PQ($T_i, W$)
**Require:** $T_i$: A term in the original query $Q$
**Require:** $W$: The term rewriting matrix
1: $M(T) \leftarrow \{\tau : W[T_i, \tau] \neq 0\}$
2: Define a priority queue $PQ_i$ (initially empty)
3: **for** each $\tau \in M(T_i)$ **do**
4:    Retrieve the first document $D$ from $\tau$
5:    Insert into $PQ_i$ a pair $(\tau, D)$ with score equal to the score of $D$ in $\tau$ multiplied by $W[T_i, \tau]$
6: **return** $PQ_i$

**Algorithm 4** GetNextDoc($PQ_i$)
**Require:** $PQ_i$: Priority queue associated with query term $T_i$
1: **if** $PQ_i$ is empty **then**
2:    **return** NULL
3: Remove the pair $(\tau, D)$ with score $s$ from the head of $PQ_i$
4: **if** $\tau$ is not exhausted **then**
5:    Retrieve next document $D'$ from list $\tau$
6:    Insert $(\tau, D')$ into $PQ_i$ with score equal to the score $D'$ in $\tau$ multiplied by $W[T_i, \tau]$
7: **return** document $D$, score $s$

THEOREM 3. *Let $P$ be the number of probes performed by the NRA algorithm when processing query $Q$ using the original uncompressed index $V$. Let $P'$ be the number of probes performed by the modified NRA algorithm when processing the same query $Q$ using a compressed index $H$ and a term rewriting matrix $W$ such that $V = WH$. The top-$k$ results returned by both algorithms are the same. Furthermore,*

$$P' \leq P + \sum_{T_i \in Q} |M(T_i)|.$$

PROOF. First, we prove that the top-$k$ documents returned by the modified NRA algorithm are the same as those returned by the unmodified NRA algorithm using the uncompressed index. Since the modified NRA algorithm differs from the unmodified NRA only in document retrieval, we only need to prove that the sequence of documents retrieved from list $L_i$ in the original index is equal to the sequence of documents retrieved from the priority queue $PQ_i$ using the compressed index through Algorithms 3 and 4.

Since $V = WH$, a row corresponding to a term $T_i$ in $V$ is equal to a linear combination of rows $M(T_i)$ in $H$, where the coefficients are in the row $T_i$ in $W$. That is,

$$\forall D \in Docs, V[T_i, D] = \sum_{\tau \in M(T_i)} W[T_i, \tau] \cdot H[\tau, D] \quad (9)$$

Since all meta-terms (i.e., rows) in $H$ corresponding to the same term in $V$ are disjoint, the value $V[T_i, D]$ can be written as $W[T_i, \tau] \cdot H[\tau, D]$, for the unique meta-term $\tau \in M(T_i)$ satisfying $H[\tau, D] \neq 0$. Algorithms 3 and 4 reconstruct the list of $T_i$ by computing the weight of each document $D$ in the list of $T_i$ as soon as $D$ appears in a list $\tau \in M(T_i)$. Moreover, the priority queue returns documents in descending order of their scores. This proves the equality of document sequences retrieved from $L_i$ and from $PQ_i$.

Now, we prove the relationship between $P$ and $P'$. Let $depth_i$ be the number of documents retrieved from $L_i$ before termination of the NRA algorithm when running on the uncompressed index.

We need to prove that the maximum number of documents retrieved from meta-term lists $M(T_i)$ is $depth_i + |M(T_i)|$. The initialization of the priority queue $PQ_i$ results in $|M(T_i)|$ probes (lines 3, 4 in Algorithm 3). Each retrieval from the priority queue results in at most one additional probe (line 5 in Algorithm 4). Thus, retrieving $depth_i$ document from priority queue requires at most $depth_i + |M(T_i)|$ probes. The total number of probes performed by the modified NRA algorithm is at most

$$\sum_{T_i \in Q} depth_i + |M(T_i)| = P + \sum_{T_i \in Q} |M(T_i)|.$$

□

Note that some meta-terms in the rewritten query might be shared across multiple terms in the original query (i.e., $M(T_i) \cap M(T_j) \neq \phi$ for $T_i, T_j \in Q$ and $i \neq j$). It is possible to further reduce the number of probes performed by the modified NRA by keeping each meta-term $\tau$ in exactly one priority queue $PQ_i$ of a term $T_i$ such that $\tau \in M(T_i)$, and removing occurrences of $\tau$ in other priority queues. Details are omitted due to space constraints.

# 6. EXPERIMENTAL RESULTS

In this section, we experimentally evaluate our approach. We evaluate the compression ratio for a representative document corpus and we show that the impact on query execution time is negligible. Finally, we investigate the effect of various parameters of the compression algorithm.

We do not directly compare our approach to other lossless compression techniques that compress posting lists individually because both approaches (i.e., holistic and per-list compression) can successfully be integrated to achieve better overall compression. To asses such fact, we show that our compression approach is nearly orthogonal to, and hence can be integrated with, a common compression technique, namely variable-byte encoding.

We perform our evaluation on memory-resident indices since this is the dominant approach in current large scale applications. The memory capacities of modern machines allow in-memory serving even from large corpora such as the entire Web [7], by partitioning the index across multiple machines. In such a setting, our technique can be applied to each index partition individually.

## 6.1 Setup

Our index factorization algorithm ran on a Hadoop cluster[1]. Query evaluation latency was measured by a single-threaded Java process running on an Intel Xeon 2.00GHz 8-core machine with 32GB RAM. Both compressed and uncompressed indices were preloaded into RAM prior to query evaluation. We used TREC WT10g document corpus [1], which contains 1.7M documents. We indexed only the textual content of the documents and discarded HTML tags. We removed the least frequent terms that appear in less than three documents, thus reducing the number of unique terms from 5.4M to 1.6M. These rare terms account for less than 1% of the index size so the effect of their removal on index compressibility is negligible. In the indices we constructed, each posting contains 4-byte integer for *docID* and 4-byte integer for payload. For query workload, we used 50,000 queries that are randomly selected from the AOL query log [21].

Unless specified otherwise, we used the following default parameter values. Block size $B$ (i.e., the maximum number of rows in a block) is set to 500. Sketch length $\lambda$ is set to 1, which means that row sketch is simply the number of non-zero elements in the row. Minimum savings threshold $\mu$ is set to 100.

[1] http://hadoop.apache.org

For measuring the compression ratio of our approach, we compute the relative reduction in the space required for storing the inverted index: (Uncompressed Index Size - Compressed Index Size) / (Uncompressed Index Size). We consider both matrices $W$ and $H$ when computing the size of a compressed index. Note that if individual posting lists are not additionally compressed by any method (e.g., var-byte encoding), the relative reduction in space is equivalent to the reduction in the total number of non-zeros in index, i.e., $\frac{\|V\|_0 - (\|W\|_0 + \|H\|_0)}{\|V\|_0}$.

## 6.2 Results

In this section, we show the results of our experiments.

**Compression Performance.** We selected two compressed indices obtained after 8 and 35 iterations of our algorithm. Figure 4 shows the compression ratio for the two indices. We observe that after 8 iterations, our factorization algorithm compresses the index by 20%, while applying var-byte encoding to the compressed index results in an overall compression of 46%. At iteration 35, the compression ratio reaches 29% and 50%, respectively. The size of matrix $W$, which maps the original terms to the meta-terms, is less than 1% of the compressed index size in all iterations.

By lowering the saving threshold $\mu$ to 0, our approach archives a compression ratio of 35% after 30 iterations (Figure 9(b)).

When limiting the number of mappers/reducers to 100 per each job, each iteration took 22 minutes in average. The runtime of the first few iterations is slightly above average (e.g., the first iteration took 27 minutes, while the second iteration took 23 minutes).

**Query Evaluation Latency.** Figure 5 shows the average query latency for different numbers of retrieved documents ($k$) using the compressed indices at iterations 8 and 35.[2] We do not show the latency of the unmodified NRA algorithm on the compressed indices as it is orders-of-magnitude higher and would distort the plot. The inefficiency of the unmodified NRA is due to the fact that the computed score bounds are very loose, which prevents early termination. Our modifications to the NRA algorithm eliminates the overhead in query evaluation and results in nearly the same performance of the unmodified NRA on the original uncompressed index. In some cases, searching a compressed index outperforms searching the uncompressed index (e.g., for $k = 20$, the latency on the index compressed by 20% is 6% lower than the latency on the uncompressed index). Thus, the optimal compression-latency tradeoff is achieved after only a few iteration.

**Size of the Factor Matrices.** Figure 6 depicts the relative number of non-zero elements in $W$ and $H$ compared to the number of non-zeros in $V$ at various iterations of the compression algorithm. Observe the monotonicity of the curve due to the property of our algorithm that never increases the number of non-zero elements in the factors. Note that matrix $W$, which is used for query rewriting, is much smaller than $H$ (e.g., $\|W\|_0$ is less than 1% of $\|V\|_0$ at iteration 35). We see that the first few iterations are the most productive, while the benefit after the tenth iteration is marginal.

**Integration with Variable-byte Encoding.** In this experiment, we show the behavior of variable-byte encoding [19] when applied to the resulting compressed index. Figure 7 shows the effectiveness of the encoding at various compression ratios of our factorization algorithm. We observe that the two techniques complement each other as they exploit different properties of the data. That is,

[2] We did not evaluate the effect of var-byte encoding on query latency, which is studied in prior works.
8

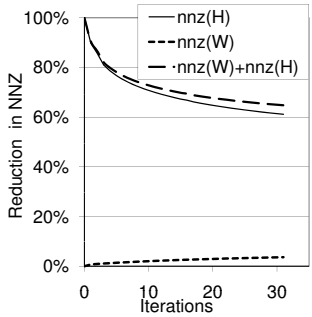
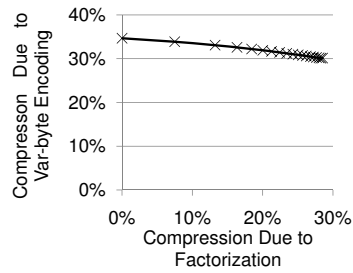
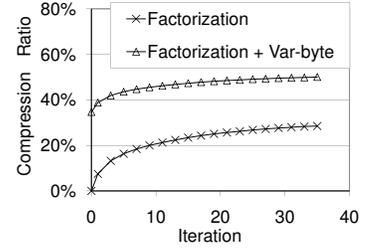

**Figure 6: Relative reduction in non-zeros**

**Figure 7: Effectiveness of var-byte encoding**

**Figure 8: Compression ratio due to factorization and var-byte**

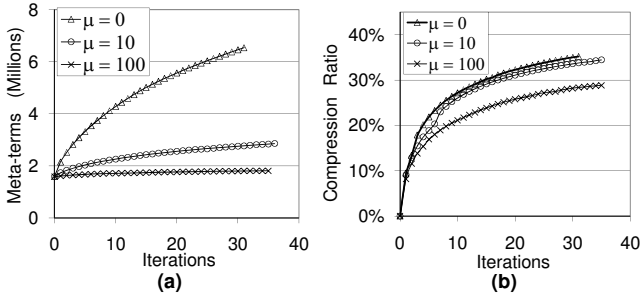

**Figure 9: The effect of $\mu$ on (a) the number of meta-terms, and (b) the compression ratio**

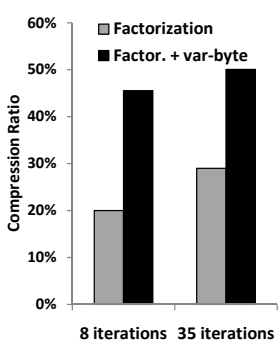
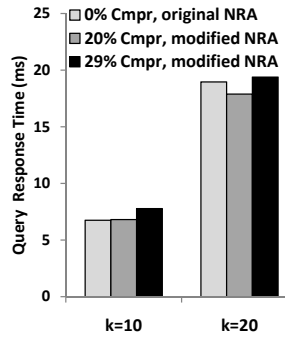

**Figure 4: Compression ratio**

**Figure 5: Average query response time**

factorization-based compression has negligible effect on the effectiveness of var-byte encoding.

Figure 8 shows the compression ratio at different iterations when using our compression method alone, and when applying the variable-byte encoding to the index generated by our factorization algorithm. The combination of the two techniques achieves compression ratio of 50% at iteration 35.

**The Saving Threshold $\mu$.** In this experiment, we analyze the effect of the savings threshold $\mu$ on the total number of meta-terms in the compressed index (Figures 9(a)), and on the compression ratio (Figures 9(b)). Recall that higher $\mu$ decreases the number of meta-terms, and thus reducing effectiveness of the compression. Changing $\mu$ from 0 to 100 reduces the total number of meta-terms in the compressed index at iteration 30 from 6.5M to 1.8M. At the same time, the compression ratio is reduced by only 6%.

Figure 10 shows the dependency between the frequency of a term in the corpus and the number of corresponding meta-terms. Clearly, the higher the term frequency, the longer the term's posting list is, resulting in more meta-terms.

**The Block Size $B$.** Figure 11 shows compression ratio for two block sizes: 10 and 500 for $\mu = 0$. When the block size is reduced by a factor of 50, the compression ratio falls by only 5%, while the average iteration runtime falls from 22 to 16 minutes. Despite the dramatic decrease in the block size, the overall runtime decreased by only 36% due to the overhead incurred by the other tasks such as row comparisons and updating $W$ and $H$, in addition to the overhead incurred by the Hadoop framework.

**The Sketch Length $\lambda$.** Figure 12 shows compression ratio for sketch lengths of 1 and 10 for $\mu = 0$. When combining terms, sketch size has no effect on the compression ratio, which means that blocking rows according to the number of non-zeros is good enough. This is due to the relatively high variability in row lengths (number of non-zeros), which is known to follow a power-law distribution. This variability provides sufficient information to identify "similar" rows.

To investigate the potential effect of sketch length, we modified our algorithm to combine columns of matrix $H$ instead of its rows (although this is not a viable option for index compression as explained in Section 3). Columns have much less variability in their lengths (number of non-zeros), since the distribution of document length is closer to normal than to power-law. In this case, blocking by column length alone is not effective, and more fine-grained similarity metrics (e.g., longer sketches) give better results (22% at iteration 7 compared to 18%).

## 7. RELATED WORK

Lossless compression of inverted indices has been an active topic for the past few years. Most of the developed techniques (e.g., variable-byte encoding, gamma-coding and delta-coding [19, 22]) aim at generating an efficient encoding of the entries in posting list, and thus can be integrated with our approach (cf. Section 6).

There are multiple techniques for lossy compression. One of the widely used techniques is *static pruning* [5, 6]. Techniques that are based on static pruning truncate postings that have low impact on the results of top-$k$ queries. The simplest form is to remove postings with payload values less than a specific cut-off threshold. Clearly, lossy compression might lead to degradation in results quality, unlike lossless compression where quality of results is not affected. In general, lossy compression can be integrated with lossless compression techniques in order to achieve higher compression ratios at the cost of lowering the quality of query results.

Several matrix factorization approaches have been proposed such as Non-negative Matrix Factorization [13, 16, 17], Principal Component Analysis [14], K-means clustering [9], Latent Seman-tic Analysis [8], and Singular Value Decomposition [23]. The goal of these techniques is to factor a given matrix into two (or three) factor matrices that (optionally) exhibit some level of sparseness. Such techniques provide a *close approximation* of the input matrix, while our approach provides an *exact* factorization of the input matrix. Modifying NMF algorithms to be lossless is not straightforward. For example, one naïve approach is to compute the remainder matrix $R = V − WH$ so that the matrix $V$ can be compactly represented using the matrices $W$, $H$, and $R$ (i.e., $V = WH + R$). Unfortunately, there is no guarantee that sparseness of $W$ and $H$ would lead to sparseness of $R$. In fact, the size of $R$ can be larger than the size of $V$ because elements in $V$ with values equal to zero may have non-zero values in the product $WH$.

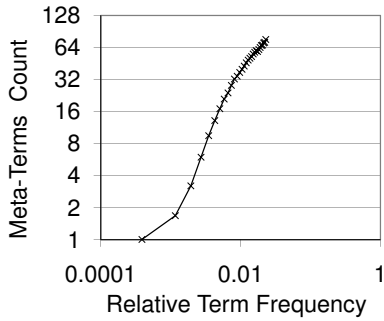
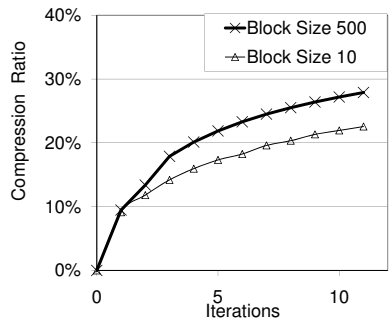
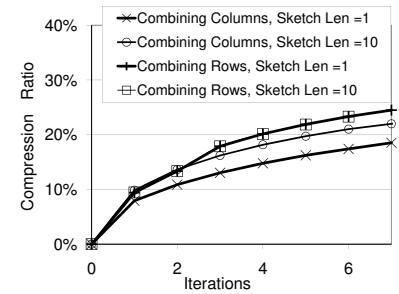

Figure 10: Number of meta-terms vs. relative term frequency in the corpus

Figure 11: The effect of the block size on compression

Figure 12: The effect of the sketch length on compression

Another line of related work in a different context, namely signal and image processing, considers a problem of representing a *signal* (vector) using a linear combination of a small number of basis vectors from a *dictionary* (e.g., [2, 24]). The problem of selecting the optimal dictionary given the set of signals is similar to the problem we consider, with two major differences: (1) the dimensions of the factor matrices are selected in advance, and (2) the sparseness is required only from the encoding vectors (matrix $W$) and not from the basis vectors (matrix $H$).

Another related problem is discovering biclusters in two-dimensional data (refer to [18] for a comprehensive survey). The goal is to discover the largest bicluster (submatrix) that exhibits certain characteristics (e.g., have the same value, or follow additive/multiplicative patterns). Computing the largest bicluster is shown to be NP-hard [18]. Our approach can be viewed as a biclustering problem (however, with a different goal) as follows. Each meta-term $\tau$ represents a bicluster in $V$ whose rows are multiples of each other and contain non-zero values only. Each bicluster results in a number of non-zero elements in matrix $W$ (respectively, $H$) that is equal to the number of rows (respectively, columns) of the bicluster. The size of a bicluster is defined as the total number of the contained rows and columns. Our goal is to obtain a set of disjoint biclusters that covers all non-zero elements of the input matrix such that the total size of biclusters is minimal. Unfortunately, previous approaches for discovering biclusters cannot be easily extended to address the described objective function.

## 8. CONCLUSION

In this paper, we presented a novel approach for compressing inverted indices without information loss. We developed a novel compression approach that is based on exact factorization of sparse matrices. We proved that obtaining the optimal factorization is NP-hard, and developed an efficient greedy factorization algorithm. We described how to modify a typical top-$k$ search algorithm to eliminate the computational overhead at the retrieval time by exploiting characteristics of our compression scheme. Our experimental evaluation shows that our technique achieves compression ratio of 35% while incurring negligible increase in the query evaluation time. We also showed that at compression ratio of 20%, the query response time is not affected by compression. Other lossless compression approaches such as variable-byte encoding can be integrated with our approach to achieve overall compression ratios up to 50%.